\documentclass[%
 reprint,
superscriptaddress,
 amsmath,amssymb,
 aps,
prb,
]{revtex4-2}

\usepackage{booktabs}
\usepackage{multirow}
\usepackage{color}
\usepackage{graphicx}
\usepackage{dcolumn}
\usepackage{bm}
\usepackage{hyperref}
\usepackage[mathlines]{lineno}
\usepackage{tabularx}

\begin{document}

\title{Multimechanism quantum anomalous Hall and Chern number tunable states in germanene (silicene, stanene)/\textit{M}Bi$_2$Te$_4$ heterostructures}

\author{Zhe Li}%
 \email{lizhe21@iphy.ac.cn}
\affiliation{%
 Beijing National Research Center for Condensed Matter Physics, and Institute of Physics, Chinese Academy of Sciences, Beijing 100190, China
}%

\author{Jiatong Zhang}
\affiliation{%
 State Key Laboratory of Low Dimensional Quantum Physics, Department of Physics, Tsinghua University, Beijing 100084, China
}%

\author{Xiyu Hong}
\affiliation{%
	State Key Laboratory of Low Dimensional Quantum Physics, Department of Physics, Tsinghua University, Beijing 100084, China
}%

\author{Xiao Feng}
\affiliation{%
 State Key Laboratory of Low Dimensional Quantum Physics, Department of Physics, Tsinghua University, Beijing 100084, China
}%
\affiliation{%
Beijing Academy of Quantum Information Sciences, Beijing 100193, China
}%
\affiliation{%
 Frontier Science Center for Quantum Information, Beijing 100084, China
}%
\affiliation{%
	Hefei National Laboratory, Hefei 230088, China
}%

\author{Ke He}
\affiliation{%
 State Key Laboratory of Low Dimensional Quantum Physics, Department of Physics, Tsinghua University, Beijing 100084, China
}%
\affiliation{%
 Beijing Academy of Quantum Information Sciences, Beijing 100193, China
}%
\affiliation{%
 Frontier Science Center for Quantum Information, Beijing 100084, China
}%
\affiliation{%
Hefei National Laboratory, Hefei 230088, China
}%

\date{\today}

\begin{abstract}
By constructing germanene (silicene, stanene)/\textit{M}Bi$_2$Te$_4$ (\textit{M} = 3$d$-transition elements) heterostructures, we’ve discovered and designed multi-mechanism quantum-anomalous-Hall (QAH) systems, including $\Gamma$-based QAH, \textit{K}-\textit{K’}-connected QAH and valley-polarized \textit{K}- or \textit{K’}-based QAH states, via first-principle computations. The unique systems possess a global gap and tunable Chern number. The coexisting conventional $\Gamma$-based QAH state of \textit{M}Bi$_2$Te$_4$ and valley-polarized \textit{K}(\textit{K’})-based QAH state of germanene (silicene, stanene), with opposite chirality, can interact with each other. Adjusting magnetic configurations of \textit{M}Bi$_2$Te$_4$-layers can not only switch on (off) the QAH conductance, but also modulate Chern numbers exactly. For example, germanene/bilayer-NiBi$_2$Te$_4$ possesses the Chern-number: \textit{C}=+1 in ferromagnetic couplings and \textit{C}=+2 in antiferromagnetic couplings. The novel multi-mechanism QAH insulators, which are achievable in experiment, provide a new approach to spintronics and valleytronics based on topological states of matter.
\end{abstract}

\maketitle

Recently, the design of magnetic topological materials towards various requirements of quantum anomalous Hall (QAH) effect attracts numerous attentions \cite{haldane1988model,yu2010quantized,chang2013experimental,weng2015quantum,mogi2015magnetic,ou2018enhancing,wang2023intrinsic,hasan2010colloquium,qi2011topological,matusalem2019quantization,matthes2016quantization,matthes2016intrinsic,li2019intrinsic,otrokov2019unique,zhang2019topological,gong2019experimental,deng2020quantum,liu2020robust,ge2020high,bai2024quantized,li2020tunable,kobialka2022dynamical,zhang2020large,tang2023intrinsic,xu2022controllable,you2019two,li2023tunable,li2020high,li2022chern,xue2024tunable,niu2019quantum,wu2023robust}. Initially put forward in certain periodically magnetic-field-assisted graphene \cite{haldane1988model} and fabricated in magnetic-doped topological insulators \cite{chang2013experimental}, QAH states inspire new research hotspots owing to its dissipationless edge-conductance, no necessity for strong external field and further investigations for Majorana fermions, axions and topological magnetoelectric fields \cite{zhang2019topological,liu2020robust,tang2023intrinsic,xu2022controllable}, etc. Noticeably, MnBi$_2$Te$_4$-family material is a representatively intrinsic van der Waals (vdW) stacked topological magnet \cite{li2019intrinsic,otrokov2019unique,zhang2019topological,gong2019experimental,deng2020quantum,liu2020robust,ge2020high,bai2024quantized,li2020tunable,kobialka2022dynamical} that has potential for both high-temperature QAH \cite{deng2020quantum,ge2020high} and Chern-number-tunable character after neighbored with monolayer Bi \cite{xue2024tunable}, in which \textit{p} electrons of MnBi$_2$Te$_4$ or Bi hybridize with \textit{d} electrons of Mn to create QAH conductance. Another route is to build QAH magnets derived from $d$-$d$ electron correlations, like Pd(Pt)Br(I)$_3$, MnBr$_3$, LiFeSe, NiAs(Bi)O$_3$ and PdSbO$_3$ \cite{you2019two,li2023tunable,li2020high,li2022chern,wu2023robust}, etc. All these systems generate QAH edge-states originated from only one certain mechanism.

Distinct to mechanisms mentioned above, the third mechanism to motivate QAH effect depends on the band-topology of \textit{K}(\textit{K’})-valleys, which is initially proposed in basic models of silicene \cite{pan2014valley,pan2015valley}. Researchers have paid great efforts to find perfect magnetic substrates to these group-IV-element monolayers \cite{zhang2015robust,zhang2018strong,zou2020intrinsic,vila2021valley,zhou2017valley,barman2023bilayer}, and predict QAH effects in some candidates \cite{zhang2015robust,zhang2018strong,zou2020intrinsic,vila2021valley,zhou2017valley}. But up to now, the predictions and accomplishments of flexibly tuning Chern and valley-polarized-Chern numbers via real materials are still lacking. Furthermore, combining and coupling multi-mechanisms of QAH in one system triggered by even one magnetic origin is still in mystery.

In this work, we systematically discover and design multi-mechanism quantum-anomalous-Hall states (mQAH) via constructing germanene (silicene, stanene)/\textit{M}Bi$_2$Te$_4$ vdW heterostructures. Among all the cases, germanene behaves the best due to its large global gaps and large $\Gamma$ (\textit{K}, \textit{K’})-based gaps. Monolayer germanene which is neighbored by MnBi$_2$Te$_4$ and NiBi$_2$Te$_4$ can generate both \textit{K}-\textit{K’}-connected QAH and valley-polarized QAH (vpQAH) states at \textit{K}- or \textit{K’}-valleys, of which total Chern number and valley-polarized-Chern number are \textit{C}=+2 and \textit{C$_v$}=-1 respectively, and the mass term is motivated by 3\textit{d} orbitals of Mn or Ni atoms. Moreover, by neighboring thicker layers of Mn(Ni)Bi$_2$Te$_4$, the mQAH state combining $\Gamma$-based QAH and \textit{K}(\textit{K’})-based vpQAH states forms, both of them triggered only by Mn(Ni)-3\textit{d} orbitals and sharing the opposite chirality. NiBi$_2$Te$_4$ behaves better than MnBi$_2$Te$_4$ with larger gaps \cite{xu2022controllable}, more appealing for experimental surveys. Remarkably, stacking-order-shift is an experimentally available method for tuning mass terms, intrinsic and extrinsic Rashba terms \cite{pan2014valley,pan2015valley} in vpQAH characters of germanene. Via stacking-order-shift, we’ve obtained phase-diagram mappings corresponding to different Chern-insulating phases, providing a paradigmatic case for switchable and tunable Chern and valley-Chern numbers in real materials.

Crystallized by low-buckled, hexagonal structure, monolayer germanene (silicene, stanene) shares similar lattice structures with \textit{M}Bi$_2$Te$_4$, in which we compare the in-plane lattice mismatch in the Supplementary Information \cite{supplementary} (see also references \cite{kresse1996efficient,perdew1996generalized,grimme2010consistent,wang2021vaspkit,togo2015first,mostofi2014updated,marzari1997maximally,souza2001maximally,wu2018wanniertools,xu2022controllable,qi2006topological,pan2014valley,kobialka2022dynamical,wang2023magnetic,guo2023novel,li2020tunable,li2019intrinsic,tang2023intrinsic} therein) as Table S1. Clearly, both germanene and stanene can approximately match \textit{M}Bi$_2$Te$_4$ by the ratio of 1:1 within $\pm$10\% of lattice mismatch, while in the case of silicene, only NiBi$_2$Te$_4$ fall in this range. The absolute value of in-plane lattice constants of these monolayer-compounds are listed in Table S2. For \textit{M}Bi$_2$Te$_4$, we select \textit{M} as Ti, V, Mn, Fe, Co, Ni due to their mechanically stable structures and large global gaps, in which the former are verified by density functional perturbation theory in Fig. S1 and the latter are confirmed by band structures (Fig. S2). Lattice mismatch brings biaxial strains into both germanene (silicene, stanene) and $M$Bi$_2$Te$_4$ without causing structural instabilities (see Figs. S3 to S6). Moreover, opting \textit{M}Bi$_2$Se$_4$ and \textit{M}Sb$_2$Te$_4$ reaches less lattice mismatch, but a rather severe requirement to achieve large-gapped QAH states (Fig. S7), so we mainly focus on that of \textit{M}Bi$_2$Te$_4$ below.

\begin{figure}
    \centering
    \includegraphics[width=1\linewidth]{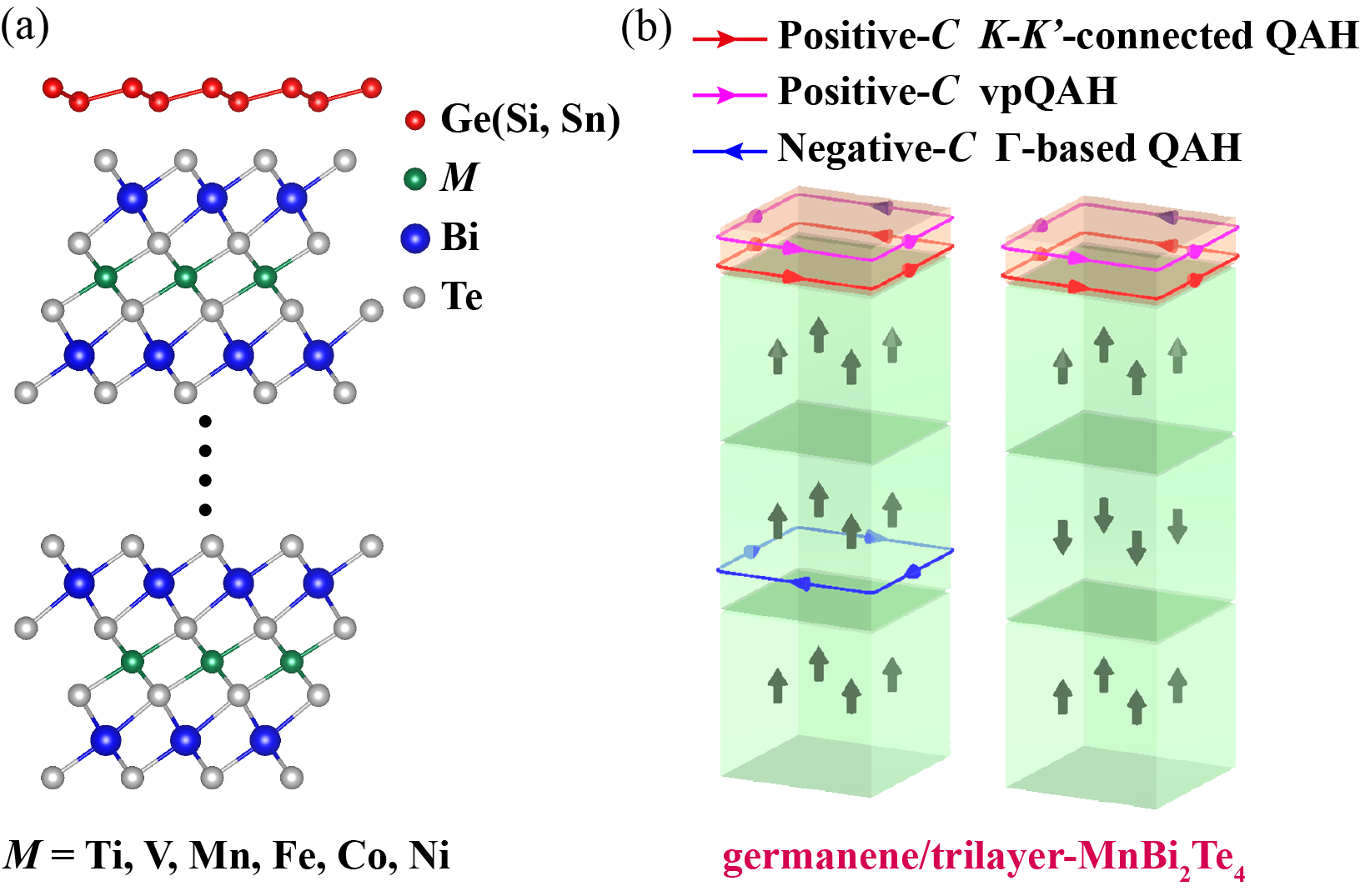}
    \caption{Lattice structures and illustrations of mQAH state in germanene (silicene, stanene)/ \textit{M}Bi$_2$Te$_4$. (a) Side view of monolayer-germanene (silicene, stanene)/few-layer-\textit{M}Bi$_2$Te$_4$, in which red, green, blue and light-gray balls denote Ge (Si, Sn), \textit{M}, Bi and Te atoms respectively. (b) Illustrations of mQAH state in germanene/trilayer-MnBi$_2$Te$_4$ under FM and AFM configurations. Light-brown and light-green blocks are layers of germanene and MnBi$_2$Te$_4$. The circular arrows colored with red, pink and blue are positive-chiralized \textit{K}-\textit{K’}-connected QAH, positive-chiralized vpQAH and negative-chiralized $\Gamma$-based QAH conductances respectively.}
    \label{fig1:Lattice}
\end{figure}

Depicted in Fig. \ref{fig1:Lattice}(a), germanene (silicene, stanene) stacks as the order “\textit{ABCABC}…” along (111) directions, same to that of multilayer \textit{M}Bi$_2$Te$_4$ itself. In this single heterostructure, \textit{M}Bi$_2$Te$_4$ itself contains conventional $\Gamma$-based QAH state under multilayer regime. In the meantime, \textit{M}Bi$_2$Te$_4$ also breaks time-reversal symmetry (TRS), brings mass terms in germanene (silicene, stanene) to create \textit{K}(\textit{K’})-based, both \textit{K}-\textit{K’}-connected QAH and vpQAH states in the latter. We call this phenomenon as multi-mechanism QAH state (abbreviated as mQAH), in which the only one origin: the \textit{M} element in \textit{M}Bi$_2$Te$_4$ triggers two or more kinds of different mechanisms of QAH states that coexist in the same system and even interact with each other. This concept unfolds a new field of QAH state, and magnetic-topology with possibly more abundant phase diagrams, compared to the mainly previous works that possess single-mechanism QAH states \cite{chang2013experimental,weng2015quantum,mogi2015magnetic,ou2018enhancing,wang2023intrinsic,li2019intrinsic,otrokov2019unique,zhang2019topological,gong2019experimental,deng2020quantum,liu2020robust,ge2020high,bai2024quantized,li2020tunable,kobialka2022dynamical,zhang2020large,tang2023intrinsic,xu2022controllable,you2019two,li2023tunable,li2020high,li2022chern,xue2024tunable,pan2014valley,pan2015valley,zhang2015robust,zhang2018strong,zou2020intrinsic,vila2021valley,zhou2017valley}.

Choosing germanene/trilayer(TL)-MnBi$_2$Te$_4$ as example, we draw illustrations of mQAH behaviors under different magnetic configurations in Fig. \ref{fig1:Lattice}(b). Ferromagnetic (FM) state MnBi$_2$Te$_4$ is necessary to generate $\Gamma$-based QAH state in itself. If the interlayer couplings of MnBi$_2$Te$_4$ recover to antiferromagnetic (AFM) state, $\Gamma$-based QAH state vanishes, meanwhile \textit{K}-\textit{K’}-connected QAH and vpQAH states still remain, protected by the most proximate MnBi$_2$Te$_4$-layer. Hereafter the computational results below will detailly describe these behaviors.

\begin{figure*}
    \centering
    \includegraphics[width=1\linewidth]{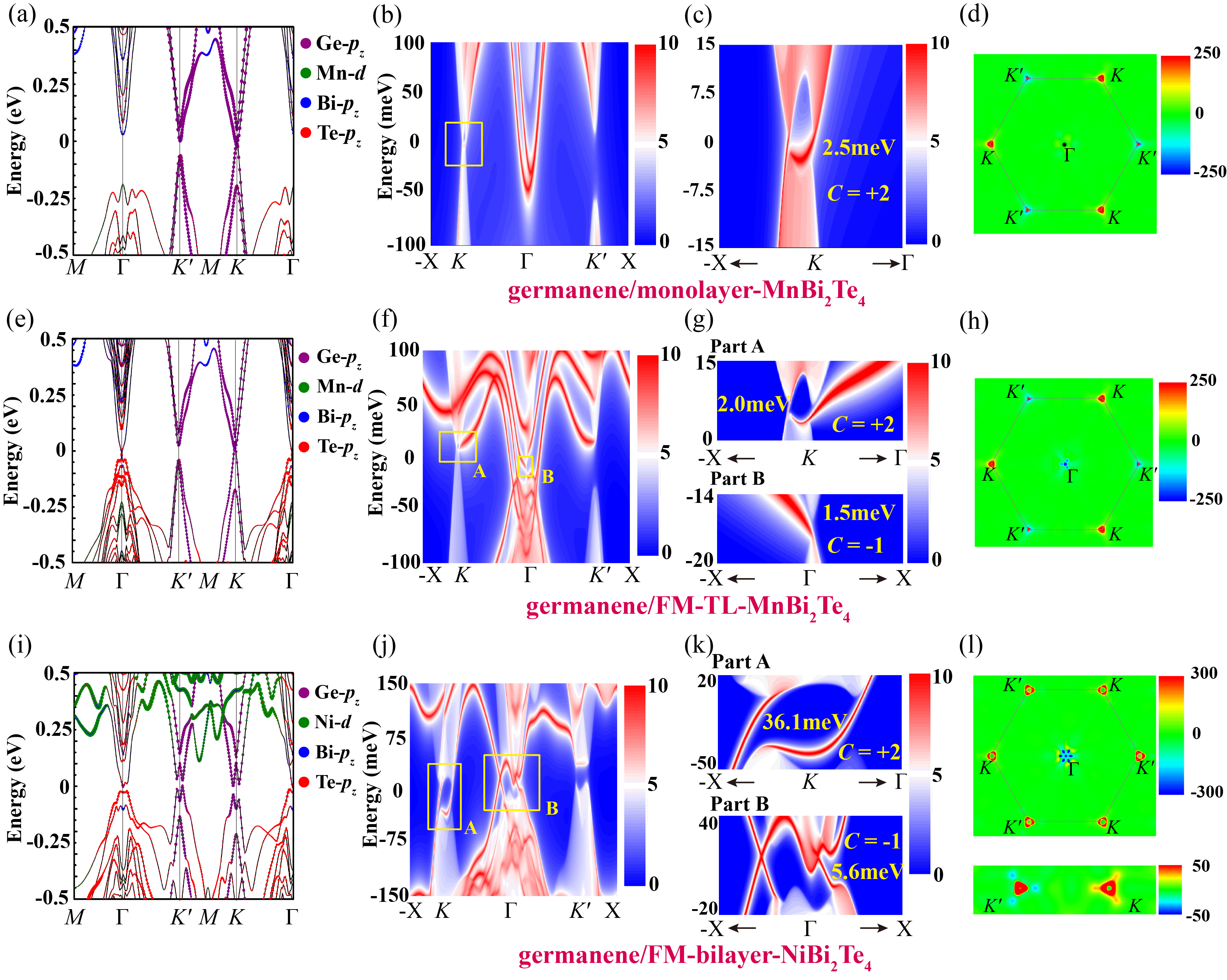}
    \caption{Band structures, LDOS and Berry curvatures of germanene/Mn(Ni)Bi$_2$Te$_4$. (a) Orbital project-band structures of germanene/monolayer-MnBi$_2$Te$_4$. Purple, olive, blue and red bubbles stand for Ge-$p_z$, Mn-$d$, Bi-$p_z$ and Te-$p_z$ contributions respectively. (b) LDOS distribution of germanene/monolayer-MnBi$_2$Te$_4$ cut along [100] boundary. (c) Zoom-in LDOS distribution around the gap at \textit{K}-valley. (d) Berry curvatures of germanene/monolayer-MnBi$_2$Te$_4$ distributing along the two-dimensional Brillouin zone. (e)-(h) are similar with (a)-(d), but in case of FM state germanene/TL-MnBi$_2$Te$_4$. (i)-(l) are also similar with (a)-(d), but under FM state germanene/bilayer-NiBi$_2$Te$_4$. In (i), olive bubbles denote Ni-$d$ contributions.}
    \label{fig2:Topology}
\end{figure*}

Figure \ref{fig2:Topology}(a) shows the orbital project-band structure of germanene/monolayer-MnBi$_2$Te$_4$. Fortunately, the gaps at $\Gamma$-point and $K$($K’$)-valley are ideally aligned to the same energy scope located at Fermi level (see Table S4). The contributions from Ge-$p_z$ orbitals mainly occupy valleys around $K$ and $K’$ points, similar to that of free-standing stanene \cite{xu2013large,xu2015large}. But the gaps around $K$-valleys are greatly shrunk compared to those around $K’$-valleys, in which the degeneracy is lifted by inversion-symmetry-broken. The asymmetry between $K$ and $K’$ is more obvious in local density-of-state (LDOS) distribution along [100] boundary [Fig. \ref{fig2:Topology}(b)]. For $K$-valley, two positive-chiralized edge states appear between conductance and valence band, meanwhile no edge state exists within the gap at $K’$-valley. Fig. \ref{fig2:Topology}(c) exactly confirm $C$=+2 character around $K$-valley, with the gap around 2.5meV. In order to verify its valley-polarized character, Berry curvature distributions are computed in Fig. \ref{fig2:Topology}(d). Obviously, Berry curvatures around $K$ and $K’$ valleys are totally different, giving integration results as +1.5 and +0.5 respectively. Therefore, the total Chern number is $C$=+2, while its valley-polarized nature is affirmed as $C_v$= $C_K’$-$C_K$ = -1. Here, a $K$-$K’$-connected QAH edge state contributing no valley-polarized nature connects valence band at $K’$-valley and conductance band at $K$-valley, exhibited in Fig. S12. MnBi$_2$Te$_4$ behaves as a multifunctional substrate via breaking both TRS and inversion symmetry, and containing strong spin-orbital coupling (SOC) effect, which leads to considerable intrinsic and extrinsic Rashba-SOC terms acting on germanene’s low-buckled structure, and global-gapped vpQAH state that was proposed with basic models \cite{pan2014valley,pan2015valley}.

Monolayer MnBi$_2$Te$_4$ itself has no QAH conductance due to its strong intralayer-quantum confinement. Considering thicker-layer conditions, and selecting FM-state TL-MnBi$_2$Te$_4$ as example, we successfully motivate $\Gamma$-based QAH state within TL-MnBi$_2$Te$_4$ while maintaining vpQAH states in germanene [Figs. \ref{fig2:Topology}(e)-\ref{fig2:Topology}(h)]. Freely-relaxed structure has band-gap-misplacement between $K$($K’$) and $\Gamma$
point, destroying global-gap. In order to obtain gap-alignment, -1.0\% of biaxial strain is implemented (see analysis in Table S4 and Fig. S8). Amazingly, the single magnetic-moment-origin of Mn element breaking TRS, produces the opposite chirality between $\Gamma$-based QAH state in TL-MnBi$_2$Te$_4$ and $K$($K’$)-based QAH state in germanene, resulting in compensated Chern number $C$=+1. No global gap exists, but regional gaps open both around $\Gamma$ point and $K$-valley, manifesting QAH edge states clearly within each gap [Fig. \ref{fig2:Topology}(g)]. Negative Berry curvature contributing the integration of -1 emerges at $\Gamma$ point relating to QAH state in TL-MnBi$_2$Te$_4$ [Fig. \ref{fig2:Topology}(h)], with valley-polarized Berry curvature retaining at $K$($K’$)-valleys, authenticating its mQAH character. Notably, when TL-MnBi$_2$Te$_4$ recovers AFM configuration, $K$($K’$)-based QAH states survive but $\Gamma$-based QAH state disappears, the total Chern number recovering to +2, evolving the Chern-tunable capability (see Figs. S13(i)-S13(k)). This behavior indicates that $K$($K’$)-based QAH states in germanene only depend on the most neighboring MnBi$_2$Te$_4$-layer. Due to weak interlayer coupling between MnBi$_2$Te$_4$ layers \cite{li2019intrinsic,li2020tunable}, a moderate strength of out-of-plane magnetic field can modulate $C$=+2 to $C$=+1, easily manipulated under experimental conditions.

NiBi$_2$Te$_4$ performs much better than MnBi$_2$Te$_4$ owing to its stronger ability to create mass terms both from Coulomb and kinetic interactions \cite{xu2022controllable}. Figs. \ref{fig2:Topology}(i)-\ref{fig2:Topology}(l) depict results of germanene/bilayer-NiBi$_2$Te$_4$ under FM coupling and biaxial strain of +1.3\%. Evidently, the global gap opens in whole Brillouin zone (BZ), with the nontrivial-gap at $K$-valley up to 36.1meV, manifesting it the potential for high-temperature QAH. Bilayer-NiBi$_2$Te$_4$ engenders not only moderate gaps at $\Gamma$-point (5.6meV) but also sizable mass term at $K$($K’$)-valleys, comprising large-gapped mQAH state. Similar to germanene/TL-MnBi$_2$Te$_4$, it generates mQAH state with $C$=+1 under FM coupling and $C$=+2 when bilayer NiBi$_2$Te$_4$ regains AFM ground state. It’s an excellent candidate for investigating $K$($K’$)-based vpQAH and $\Gamma$-based QAH states, but hard for Chern-tunable applications due to its stronger interlayer magnetic couplings \cite{li2019intrinsic,li2020tunable}.

We systematically investigate germanene/$M$Bi$_2$Te$_4$ when $M$=Ti, V, Mn, Fe, Co, Ni by setting out-of-plane magnetism firstly. Among all of them, Ti and V has no ability to induce QAH conductance limited by their weak Coulomb and kinetic interactions \cite{xu2022controllable}. Mn holds enough ability to motivate vpQAH states in germanene from monolayer to TL, but possess mQAH state only at FM, TL regime. Fe and Co behave similar with that of Mn, but its potential mQAH state is blocked by band-overlapping in TL (Figs. S16 and S17). Chern numbers of the only two candidates of mQAH and Chern-number-tunable systems (germanene/TL-MnBi$_2$Te$_4$ and germanene/bilayer-NiBi$_2$Te$_4$) are listed in Table \ref{tab1:tunable_Chern} with red, and the LDOS results of all the discussed heterostructures above are depicted in Fig. \ref{fig2:Topology} and Figs. S13-S17, verifying Chern numbers in every condition listed in Table \ref{tab1:tunable_Chern}. Rule out the other four candidates, we list magnetic crystalline anisotropy energies of germanene/TL-MnBi$_2$Te$_4$ and germanene/bilayer-NiBi$_2$Te$_4$ in Table S6, confirming their ground state lying in out-of-plane magnetism. 

\begin{table}
\caption{\label{tab1:tunable_Chern}Chern-insulating and tunable behaviors of germanene/$M$Bi$_2$Te$_4$. Red digits are denoted as Chern-number-tunable conditions, while “Metallic” means the system has no gap whether at $\Gamma$ point or at $K$($K’$)-valley. All the cases are derived under the out-of-plane magnetic-moment condition.}
\begin{ruledtabular}
    \begin{tabular}{ccccc}
        \multicolumn{1}{c}{$M$Bi$_2$Te$_4$} &
          \multicolumn{1}{c}{Magnetism} &
          \multicolumn{1}{c}{Monolayer} &
          \multicolumn{1}{c}{Bilayer} &
          \multicolumn{1}{c}{Trilayer} \\ \hline
        \multirow{2}{*}{TiBi$_2$Te$_4$} & FM  & 0   &   0   & Metallic   \\
        & AFM & - &    0   & Metallic    \\ \hline
        
        \multirow{2}{*}{VBi$_2$Te$_4$} & FM  & 0   &    0   & Metallic   \\
        & AFM & - &    0   & Metallic    \\ \hline
        
        \multirow{2}{*}{MnBi$_2$Te$_4$} & FM  & +2   &   +2   & \textcolor{red}{+1}   \\
        & AFM & - &    +2   & \textcolor{red}{+2}    \\ \hline
        
        \multirow{2}{*}{FeBi$_2$Te$_4$} & FM  & +2   &    +2    & Metallic   \\
        & AFM & - &   +2   & Metallic    \\ \hline
        
        \multirow{2}{*}{CoBi$_2$Te$_4$} & FM  & +2   &    +2    & Metallic   \\
        & AFM & -  &    +2   & Metallic    \\ \hline
        
        \multirow{2}{*}{NiBi$_2$Te$_4$} & FM  & +2   &    \textcolor{red}{+1}    & Metallic   \\
        & AFM & -  &    \textcolor{red}{+2}   & Metallic    \\ 
        \end{tabular}
\end{ruledtabular}
\end{table}

It's worth to mention that moderate biaxial strains cause no influence on the topological features expect the case that located very near to the phase transition point (FM state germanene/TL-MnBi$_2$Te$_4$), shown in Figs. S9 and S10. Furthermore, increasing the values of Hubbard $U$ also fails to induce topological phase transitions with only local band-gap decreasing, exhibited in Fig. S11. These outcomings verify the robustness of these novel magnetic topological characters in most cases.

In view of vdW-stacking nature of germanene and $M$Bi$_2$Te$_4$, conveniently stacking-order-shift is an exercisable way to experimentally and continuously manipulate the interplay between germanene and $M$Bi$_2$Te$_4$. Thereinafter, we focus on Chern-number-tunable phase transitions, global or regional gap-modulating outcomings relying on stacking-order-shifts.

\begin{figure*}
    \centering
    \includegraphics[width=1\linewidth]{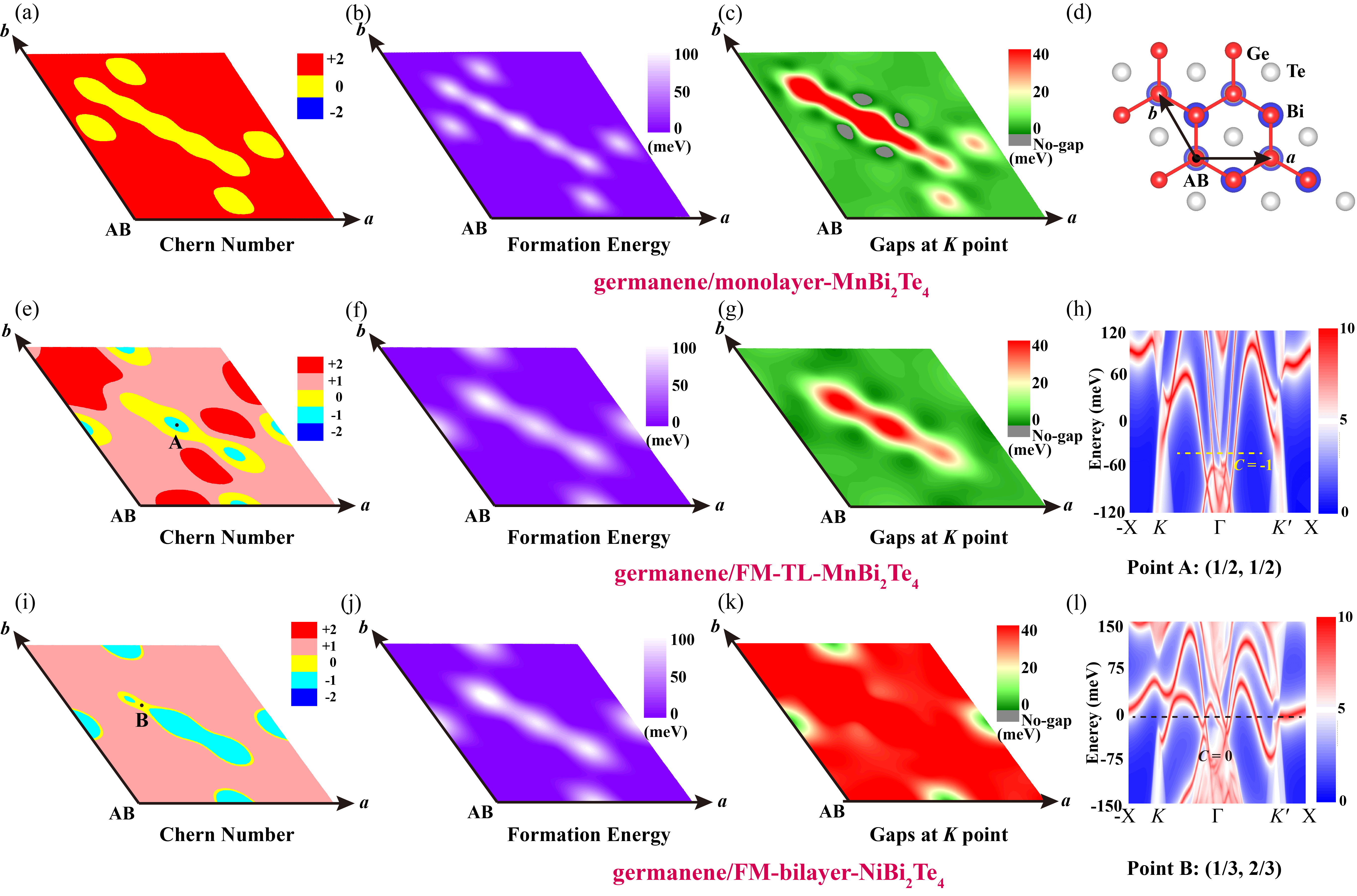}
    \caption{Stacking-order-shift investigations in germanene/MnBi$_2$Te$_4$ under monolayer- and TL-MnBi$_2$Te$_4$ cases. (a)-(c) are the case of germanene/monolayer-MnBi$_2$Te$_4$. (a) Chern-number distributions along one in-plane shift primitive cell. Red and yellow regions are corresponding to $C$=+2 and $C$=0 phase respectively. "AB" in the original point means the normal stacking order. (b) The formation energy distributions by setting that under the normal stacking order to zero. From purple to white, the formation energy increases. (c) The gap-distributions of $K$-valley. From green to yellow and red, the gap increases. Dark-gray region means no gap. (d) A schematic illustration of stacking-order-shifts between germanene and $M$Bi$_2$Te$_4$. Red, light-gray and blue balls stand for Ge, Te and Bi atoms. Two black arrows are related to primitive axes on in-plane shift. (e)-(g) are similar to (a)-(c), but under FM state germanene/TL-MnBi$_2$Te$_4$. (h) LDOS pattern along [100] boundary at the point A (1/2, 1/2) of stacking order noted in (e). (i)-(k) are similar to (a)-(c), but under FM state germanene/bilayer-NiBi$_2$Te$_4$. (l) LDOS pattern along [100] boundary at the point B (1/3, 2/3) of stacking order noted in (i).}
    \label{fig3:Stackings}
\end{figure*}

Figure \ref{fig3:Stackings} displays and analyzes stacking-order-induced phase mappings of germanene/monolayer-MnBi$_2$Te$_4$, FM state germanene/TL-MnBi$_2$Te$_4$ and FM state germanene/bilayer-NiBi$_2$Te$_4$ that discussed in Fig. \ref{fig2:Topology}. For all the three cases, Chern numbers can be step-likely and flexibly modulated with different stacking orders. Considering the case germanene/monolayer-MnBi$_2$Te$_4$ displayed in Figs. \ref{fig3:Stackings}(a)-\ref{fig3:Stackings}(c), in which Chern numbers, forming energies and the gap at $K$-valley distribute along in-plane-shift primitive cell are shown as mapping patterns. Fig. \ref{fig3:Stackings}(d) illustrates concretely how does germanene-layer shifts on the $M$Bi$_2$Te$_4$ layer, within which two black arrows noted as \textit{\textbf{a}} and \textit{\textbf{b}} mark the two primitive axes. The original point named as “AB” corresponds to normal “$ABCABC$…” stacking order. After atomical relaxations, different stacking-orders mainly fall into two phase categories, related to major region of $C$=+2 [red zone in Fig. \ref{fig3:Stackings}(a)] and minor region of $C$=0 [yellow zone in Fig. \ref{fig3:Stackings}(a)] respectively. These two distinguished phase distributions are directly related to low and high formation energy regions in Fig. \ref{fig3:Stackings}(b), corresponding to small ($\sim$2.7Å) and large ($\sim$3.0Å) vdW distances respectively (see Fig. S18). Intuitively, larger vdW distance, and higher formation-energy weakens electronic hybridization between germanene and MnBi$_2$Te$_4$, reducing both the intrinsic, extrinsic Rashba-SOC terms and the mass terms, then eliminating both the $K$-$K’$-connected QAH and vpQAH states in germanene. See in Fig. \ref{fig3:Stackings}(c), the gaps around $K$-valley also fall into two categories, in which the smaller one (2meV$\sim$4meV) corresponding to $C$=+2 condition while the larger one (above 40meV) related to topologically-trivial property. Although the $K$-valley gaps are small and not sensitive to stacking-order-shift, in most regions the whole system holds BZ-global gaps (see Fig. S18), profitably for experimental investigations and applications. 

The condition when germanene is proximate to TL-MnBi$_2$Te$_4$ under FM coupling looks more complicated. Both the Chern number of germanene and TL-MnBi$_2$Te$_4$ can be modulated with stacking-order reformation (see Fig. S19). Fig. \ref{fig3:Stackings}(e) delineates a more abundant Chern-number-distribution mapping, offering a step-likely Chern-tunable method from +2 to -1. Similarly, the lower Chern number (-1 and 0) region is related to white-colored, higher formation energies in Fig. \ref{fig3:Stackings}(f), in which the Chern number in germanene is zero (see Fig. S19). For the higher-Chern number (+1 and +2) region, the $K$-valley gaps retain between 2meV to 5meV, similar to the monolayer-MnBi$_2$Te$_4$ case. The light-red area of $C$=+1 occupies the most part, affirming that mQAH state is active under most stacking-orders. Regrettably, no BZ-global gap exists totally in the whole shifting plane [Fig. S18(g)]. We select the shifting position: (1/2, 1/2) that fall in the center of Fig. \ref{fig3:Stackings}(e), signed as “Point A”, which holds Chern number as -1, to search more details. Fig. \ref{fig3:Stackings}(h) exhibits the LDOS of Point A, in which no chiral edge states cross through $K$- and $K’$-valley gaps, but a residual negative-chiralized edge state crosses through the gap at $\Gamma$ point, revealing that the $C$=-1 is rooted in TL-MnBi$_2$Te$_4$ itself. 

Moreover, we depict the stacking-order-shift behaviors of FM-state germanene/bilayer-NiBi$_2$Te$_4$ in Figs. \ref{fig3:Stackings}(i)-(l), which shares similar performance	 with that of germanene/FM-TL-MnBi$_2$Te$_4$, but possesses rather more robust Chern-insulating property orginated from bilayer-NiBi$_2$Te$_4$, excluding $C$=+2 area absolutely [Fig. \ref{fig3:Stackings}(i)].  Remarkably, Ni contributes larger mass terms than Mn \cite{xu2022controllable}, bringing and retaining larger $K$-valley gaps even above 30meV at the most part of stacking-order-shift positions [Figs. \ref{fig3:Stackings} (j) and \ref{fig3:Stackings}(k)]. Fig. \ref{fig3:Stackings}(l) shows the LDOS pattern in the position (1/3, 2/3) shown in Fig. \ref{fig3:Stackings}(i) that labeled as "Point B", in which the "$C = 0$" nature is originated from a competitive character of Chern-insulating mechanisms from opposite chirality. Consequently, stacking-order-shifts can modulate Chern numbers step-likely and glibly meanwhile persisting large-gap characters based on this system.

\begin{figure*}
	\centering
	\includegraphics[width=1\linewidth]{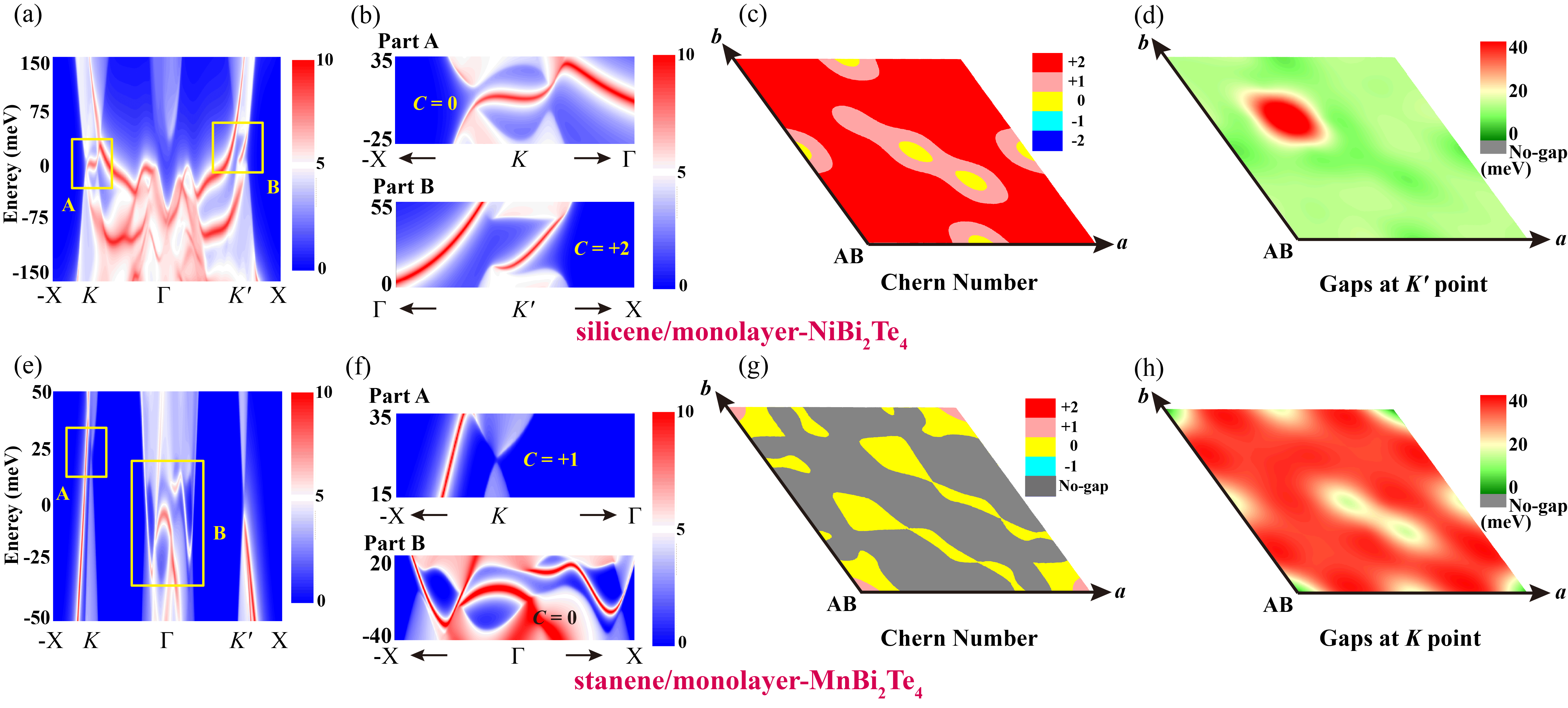}
	\caption{LDOS of normal stacking-order, stacking-order-shift investigations of Chern numbers and $K$($K’$)-valley gaps in silicene/monolayer-NiBi$_2$Te$_4$ and stanene/monolayer-MnBi$_2$Te$_4$. (a)-(d) are under the case of silicene/monolayer-NiBi$_2$Te$_4$. (a) LDOS patterns of [100] boundary under normal “$ABCABC$…” stacking-order. (b) Two zoom-in LDOS patterns of (a) labeled as “Part A” and “Part B” respectively, marked with yellow frame in (a). (c) Stacking-shift-dependent Chern-number distributions within in-plane-shift primitive cell. Red, light-red, yellow, cyan and blue regions are related to $C$ = +2, +1, 0, -1, -2 respectively. (d) Chern-insulating $K’$-valley gap distributions within in-plane-shift primitive cell. From green to yellow and red the gap increases. (e)-(h) are analogous with that of (a)-(d), but correspond to the case of stanene/monolayer-MnBi$_2$Te$_4$. In (h), we choose nontrivial $K$-valley gaps, not trivial $K’$-valley gaps.}
	\label{fig4:Stackings-SnSi}
\end{figure*}

Owning the same group element and analogous structure, silicene and stanene are expected to possesses mQAH state including vpQAH state after neighboring $M$Bi$_2$Te$_4$. Figure \ref{fig4:Stackings-SnSi} takes silicene/monolayer-NiBi$_2$Te$_4$ and stanene/monolayer-MnBi$_2$Te$_4$ as two instances. We choose +2.0\% and +2.6\% biaxial strains in order to align $\Gamma$-gap and $K$($K’$)-gap to similar energy-scope for silicene/monolayer-NiBi$_2$Te$_4$ and stanene/monolayer-MnBi$_2$Te$_4$ respectively. Opposite to that of germanene/Mn(Ni)Bi$_2$Te$_4$, for silicene/monolayer-NiBi$_2$Te$_4$ the Chern-insulating phenomenon is located on the gap of $K’$-valley, not $K$-valley. Zoom-in LDOS patterns in Fig. \ref{fig4:Stackings-SnSi}(b) uncover the details, within which no valley-polarized nontrivial edge-state exists at $K$-valley (see “Part A”), meanwhile one $\Gamma$-$K’$-connected QAH edge-state emerges, and another vpQAH edge-state appears at $K’$-valley, see “Part B” in Fig. \ref{fig4:Stackings-SnSi}(b). Immediately, we arrive the conclusion that the total Chern number $C$=+2 and the valley-polarized Chern number $C_v$=+1. Stacking-order-shift will modulate the Chern number as +2, +1 and 0 [Fig. \ref{fig4:Stackings-SnSi}(c)] while maintaining the gap at $K’$-valley to about 15meV$\sim$20meV in most positions [Fig. \ref{fig4:Stackings-SnSi}(d)]. Surprisingly, in the small zone around the position (1/3, 2/3), $K’$-valley gap jumps to above 40meV with nontrivial Chern-insulating property in the meantime. In this system, not only the total Chern number but also the valley-polarized Chern number can be step-likely modulated, shown in Fig. S20. $C_v$ is tuned as +1, 0, -1, the function of which is absent in germanene/Mn(Ni)Bi$_2$Te$_4$ (see Fig. S18). This is the first predicted valley-polarized Chern-number-tunable material that is experimentally executable via only stacking-shifts. Detailly, the valley-polarized Chern number distributions of silicene/monolayer-NiBi$_2$Te$_4$ show no correspondence with formation energies [Figs. S20(a) and S20(b)]. The global gaps of it vanish in most areas, but exist in several areas smaller than 15meV [Fig. S20(c)]. The orbital-projected band structures of silicene/monolayer-NiBi$_2$Te$_4$ and stanene/monolayer-MnBi$_2$Te$_4$ are displayed in Fig. S21.

Stanene behaves poorly compared to that of germanene and silicene on account of its energy-unbalance between the gap at the $\Gamma$ point and the $K$($K’$)-valley when choosing MnBi$_2$Te$_4$ as substrate \cite{barman2023bilayer}. For stanene/monolayer-MnBi$_2$Te$_4$, computational outcomings support a small gap at both $\Gamma$ and $K$($K’$) point, which is further confirmed in LDOS patterns [Fig. \ref{fig4:Stackings-SnSi}(e)]. From zoom-in parts in Fig. \ref{fig4:Stackings-SnSi}(f), $K$-valley contributes $C$=+1 and $\Gamma$ point supplies $C$=0, manifesting it a $K$-$K’$-connected QAH insulator with total Chern number $C$=+1. Stacking-order-shift now acts as collapsing force to Chern-insulating character, destroy its QAH conductance to trivial feature that labeled as yellow zone, and even metallic phase that labeled as dark-gray region in Fig. \ref{fig4:Stackings-SnSi}(g). After shifting stanene on the MnBi$_2$Te$_4$, the gap on the $K$-valley expands above 30meV for most positions, greatly higher than normal stacking order that is only about 3meV [Fig. \ref{fig4:Stackings-SnSi}(h)]. Stanene/MnBi$_2$Te$_4$ has been successfully experimentally-fabricated \cite{barman2023bilayer} assisted with experiences of growing stanene on Bi$_2$Te$_3$ \cite{zhu2015epitaxial}, therefore it’s the most probable candidate that achieves $K$-valley QAH conductance under transport measurements, let alone its barely satisfactory behaviors. Silicene and stanene perform badly for their metallic features as $M$Bi$_2$Te$_4$ grows thicker, blocked the potential for accomplishing and measuring mQAH states (see Figs. S22 and S23).

In summary, we discover and carefully burrow the mQAH characters simply via constructing germanene (silicene, stanene)/$M$Bi$_2$Te$_4$ heterostructures. Among these systems, germanene is induced as both $K$-$K’$-connected QAH and vpQAH states as long as $M$ selected as Mn, Fe, Co and Ni. Under FM couplings, germanene/TL-MnBi$_2$Te$_4$ and germanene/bilayer-NiBi$_2$Te$_4$ trigger $K$($K’$)-based QAH and $\Gamma$-based QAH state originated from only $M$ elements, producing the opposite chirality between germanene and $M$Bi$_2$Te$_4$, causes compensated total Chern number. For the condition of bilayer-NiBi$_2$Te$_4$, the BZ-global gap survives in the most stacking-orders, and the $K$-valley gap remains above 30meV towards high-temperature applications. Stacking-order-shifts produce a step-likely Chern-number-tunable method utilizing these heterostructures, for which the Chern number is flexibly modulated among +2, +1, 0, -1 meanwhile retaining the size of $K$-valley gap and the $\Gamma$ gap. For the first time, we not only put forward the concept of mQAH state which is predicted in some candidates of germanene/$M$Bi$_2$Te$_4$, but also establish a paradigmatic materials’ scheme for tunable total-Chern and valley-polarized-Chern insulator, greatly beneficial for spintronic (valleytronic) applications.

\begin{acknowledgments}
	We thank for X.-Y. Tang and Y. Chen for helpful discussions. This work was supported by the National Natural Science Foundation of China (92065206).
\end{acknowledgments}

\newpage
\nocite{*}
\bibliography{Main_text}

\end{document}